\documentclass[a4paper,12pt]{article}
\usepackage{epsfig}
\advance \topmargin by -\headheight
\advance \topmargin by -\headsep
\evensidemargin \oddsidemargin
\advance\hoffset by -3mm  % A4 is narrower.
\def\doubleline#1{\overline{\overline{#1}}}

\def\underDownarrow{\mathrel{\mathop{\downarrow}\limits_{-}}}
\def\underUparrow{\mathrel{\mathop{\Uparrow}\limits_{-}}}
\def\underdownarrow{\mathrel{\mathop{\downarrow}\limits_{-}}}
\def\underuparrow{\mathrel{\mathop{\uparrow}\limits_{-}}}

\def\uparrowA{\mathrel{\mathop{\uparrow}\limits_{\mathrm{A}}}}

\def\downarrowA{\mathrel{\mathop{\downarrow}\limits_{\mathrm{A}}}}
\def\DownarrowA{\mathrel{\mathop{\Downarrow}\limits_{\mathrm{A}}}}
\def\uparrowB{\mathrel{\mathop{\uparrow}\limits_{\mathrm{B}}}}
\def\UparrowB{\mathrel{\mathop{\Uparrow}\limits_{\mathrm{B}}}}
\def\downarrowB{\mathrel{\mathop{\downarrow}\limits_{\mathrm{B}}}}

\def\overDownarrow{\mathrel{\mathop{\downarrow}\limits^{-}}}
\def\overUparrow{\mathrel{\mathop{\Uparrow}\limits^{-}}}

\begin{document}

{\title{A model for ensemble NMR quantum computer using antiferromagnetic
structure}}

{\author{A. A. Kokin}}
\date{}

\maketitle 

\par{Institute of Physics and Technology, Russian Academy of Sciences,
Nakhimovskii pr. 34, Moscow 117218, Russia, E-mail:
kokin@poboxes.com}

\begin{abstract}
The one-dimensional homonuclear periodic array of nuclear spins 
$I = 1/2$, owing to hyperfine interaction of nuclear spins with
electronic magnetic moments in antiferromagnetic structure, is
considered. The neighbor nuclear spins in such array are opposite
oriented and have resonant frequencies determined by hyperfine
interaction constant, applied magnetic field value and interaction
with the left and right nuclear neighbor spins. The resonant
frequencies difference of nuclear spins, when the neighbor spins
have different and the same states, is used to control the spin
dynamics by means of selective resonant RF-pulses both for single
nuclear spins and for ensemble of nuclear spins with the same
resonant frequency.\par A model for the NMR quantum computer of
cellular-automata type based on an one-dimensional homonuclear
periodic array of spins is proposed. This model may be generalized
to a large ensemble of parallel working one-dimensional arrays and
to two-dimensional and three-dimensional structures.
\end{abstract}
\par
\section*{Introduction}
\par
The fundamental obstacle, preventing experimentalists from
extending the number of qubits to $L \gg  1$ in an individual
molecule of the liquid-state NMR quantum computer, is
the difficulty of distinguishing $L$ unique set of two-state
cells. To remove this obstacle it was already proposed several
models for solid-state quantum computers with both {\it individual}
and {\it ensemble} control of qubits. One of such
potentially realizable model based on a one-dimensional cellular
automaton, using an one-dimensional periodic array ABCABC\ldots\ of
{\it three types} of two-state quantum-mechanical cells (they may
be heteronuclear system of spins $I = 1/2$) with distinct resonant
frequencies and local interaction between near neighbors, was
first considered by S.Lloyd \cite{1}. The effect of the
interaction contains in a shift of the each cell energy levels
depending on states of its neighbors. After using the resonant
$\pi -$pulse all cells of type A, for instance, invert their state
if, and only if, the left neighbor C is in ground state and the B
on its right is in excited state. In \cite{1}, it was represented
algorithm, which was applied globally to all cells, so that there
is no need to address cells individually. This model was recently
developed by S.Lloyd \cite{17}.\par The more general model of a
solid state {\it ensemble} NMR quantum computer was described in
\cite{2}, where it was considered periodic structure of
ABCABCABC\ldots\ type in two or three dimensions with the nuclear
spins 1/2 {\it only of three} distinguish types A, B, C. It was
supposed that the nuclei are embedded in a crystal lattice of some
solid state compound with spinless nuclei and all spins are
initialized to the ground state $|0\rangle $. Each ABC-unit of
this superlattice can be used to store quantum information by
setting one of spin up or down. This information can be moved
around via some quantum cellular state shifting mechanism.
Cascading {\it unitary quantum} SWAP operations of
$A\Leftrightarrow \mathrm{B}$, $B\Leftrightarrow \mathrm{C}$,
$C\Leftrightarrow \mathrm{A}$, $A\Leftrightarrow \mathrm{B}$,\ldots\ 
is used for this process. An ancillary dopand
nucleus $D$ with spin 1/2 in the proximity of an A-site can serve
as the input/output port. A local environment region near dopand
nucleus provides a large quantum system with a wealth of qubits
and only three types of nuclear spins. Therefore impurity doping
may induce large-scale quantum automata in a single crystal and
the whole crystal contains a {\it huge ensemble} of such identical
NMR quantum computers --- large artificial "molecules".\par
One-dimensional scheme that was based {\it only on two} different
A and B types of cell in a periodic array without the ability to
distinguish the left neighbor from the right was described in
\cite{3}. Each two-state cell of the scheme has ground
$|\downarrow \rangle$ and excited $|\uparrow \rangle$ internal
eigenstates and can represent any quantum superposition of these
states. All cells are initially in the same ground states
$|\downarrow \rangle$ and the state of the all array is
$|\downarrowA \downarrowB \downarrowA\ldots \downarrow \rangle$,
similar to an one-dimensional two-sublattice ferromagnetic. Each
qubit of quantum information in the state is represented by four
consecutive units: the qubit basis state "0" is represented by
unit $|\uparrowA \uparrowB \downarrowA \downarrowB\rangle $,
whilst the state "1" is represented by $|\downarrowA \downarrowB
\uparrowA \uparrowB\rangle $.\par The model of array described
below could be realized by using a linear artificial "molecule"
with A and B cells alternating along its length in
antiferromagnetic-type structure. As the cells in this array are
used {\it only identical} nuclear spins $I = 1/2$. The neighbor
nuclear spins in the ground state of antiferromagnetic structure
are {\it opposite} orientated and have distinct resonant
frequencies determined by hyperfine interaction constant, by
applied magnetic field value and by interaction with the left and
right nuclear neighbor spins. The major advantage of this variant
over the ferromagnetic structure is that the antiferromagnet
doesn't have the total spontaneous magnetization and the nuclear
resonance frequency doesn't depend on the sample shape.\par
\par
\section{The one-dimensional antiferromagnetic model \sloppy on atoms $^{31}\mathbf{P}$.}
\par
In \cite{4,5} it was suggested a bulk-ensemble generalization of
the silicon quantum computer model proposed by Kane previously
\cite{6}. In ensemble case, unlike the individual Kane's model, 
two-type electrodes $\mathbf{A}$ and $\mathbf{J}$ form a set of narrow 
$(l_{\mathrm{A}} \sim  10\,\mathrm{nm})$ and long
(several micrometers) strips. The distance between neighbors $\mathbf{A}$
gates was assumed $l_{\mathrm{x}} \sim  l_{\mathrm{A}}$. Along the gates $\mathbf{A}$, donor $^{31}\mathrm{P}$ atoms $l_{\mathrm{y}}$
distant from each other are placed. If exchange interaction
constant for localized electronic spins along the strip gates is
more than for electronic spins between neighboring strips and more
than Zeeman energy $J(l_{\mathrm{y}}) \gg  J(l_{\mathrm{x}}), 2\mu _{\mathrm{B}}B$ ($B$ is the induction of the
applied magnetic field), it produces an artificial one-dimensional
{\it antiferromagnetically ordered state} of electronic spins. At the
temperatures well below the critical temperature (Neel
temperature) $T_{\mathrm{N}\mathrm{S}} \sim  J(l_{\mathrm{y}})/k$ ($k$ --- the Boltzmann constant) we will
have {\it a pure macroscopic} electronic ground quantum state. Due to
hyperfine interaction nuclear spins will be oriented according to
the electronic spin direction in the resultant field and will form
array with the alternating orientation of nuclear spins. Note,
this state is not the true pure nuclear antiferromagnetic state so
as long as the phases of distinct nuclear spins at macroscopic
distances are not correlated at temperature of order or higher
then critical temperature of nuclear magnetic dipole ordering,
that is $T > T_{\mathrm{N}\mathrm{I}} \sim  (10^{-6} - 10^{-7}) K$ \cite{7}. 
However, the phase correlations of near neighbor nuclear spins of course exist.\par
The nuclear resonant frequencies $\nu _{\mathrm{A},\mathrm{B}}$ of neighbor nuclear
spins are different for each of the magnetic quasi-one-dimensional
subarrays A and B in the chain and depend on the states of
neighboring spins. We will take it in the form:
\begin{eqnarray}
\nu _{\mathrm{A},\mathrm{B}}(m_{<} + m_{>}) \approx  |g_{\mathrm{N}}\mu _{\mathrm{N}}B \pm  A/2 - I_{\mathrm{n}}(m_{<} + m_{>})|/2\pi \hbar  ,\label{1}
\end{eqnarray}
where $\mu _{\mathrm{N}} = 5.05\cdot 10^{-27}\,\mathrm{J}/\mathrm{T}$ is the nuclear magneton, $A$ is hyperfine
interaction constant, (for $^{31}\mathrm{P}$: $g_{\mathrm{N}} = 2.26$, $A = 7.76\cdot 10^{-26}\,\mathrm{J})$, $I_{\mathrm{n}}$ ---
the constant of two neighbor nuclear indirect spin-spin
interaction, $m_{<}$ and $m_{>}$ are the magnetic quantum numbers for the
left and right spins. The nonsecular part of interaction is
neglected here taking in to account that $g_{\mathrm{N}}\mu _{\mathrm{N}}B$, $A/2 \gg  I_{\mathrm{n}}$.\par
Thus we have the one-dimensional homonuclear periodic array of
nuclear spins $I = 1/2$, formed in the one-dimensional
antiferromagnet at the applied magnetic field, owing to hyperfine
interaction of nuclear spins with the electronic magnetic moments.
At magnetic fields $B \geq  A/2g_{\mathrm{N}}\mu _{\mathrm{N}} \sim  3.5\,\mathrm{T}$ and at temperatures
$T \sim  10^{-3}\,\mathrm{K}$ the nuclear spins $^{31}\mathrm{P}$ have in each subarray almost 100\%
orientation (2$\pi \hbar \nu _{\mathrm{A},\mathrm{B}}/kT \leq  1)$, that is they are {\it in ground} state.
Note, that the using of dynamic methods, such as optical pumping,
makes possible the high orientation of nuclear spins also at more
large temperatures.\par
We will estimate here the exchange interaction constant
$J > 2\mu _{\mathrm{B}}B \sim  6.5\cdot 10^{-23}\,\mathrm{J}$, the critical temperature $T_{\mathrm{N}\mathrm{S}} \sim  J/k \sim  4.5\,\mathrm{K}$
and the nuclear spin critical temperature, that is due mainly to
the Suhl-Nakamura indirect spin-spin interaction,
$T_{\mathrm{N}\mathrm{I}} \sim  I_{\mathrm{n}}/k \sim  A^{2}/Jk \sim  10^{-5}\,\mathrm{K}$.\par
Here we shall use for the organization of logic operations the
addressing to spin states and qubits, analogously to consideration
\cite{3}. We shall consider at first the simple one-dimensional model
of the antiferromagnet, in which each cell is represented by the
magnetic atom and has one electronic and one nuclear spin 1/2 with
hyperfine interaction, similar to the mentioned above artificial
molecule of antiferromagnetically ordered donors $^{31}P$ in silicon
substrate.\par
Nuclear spins of identical atoms at $g_{\mathrm{N}}\mu _{\mathrm{N}}B < A/2$ are oriented
according to the electronic spin direction in the resultant field
and will form a periodic ground state array of ABAB\ldots\ type:
$\uparrow  \downarrow  \uparrow  \downarrow  \ldots $, where $\uparrow $ marks the ground state of nuclear spin in an
A-site and $\downarrow $ --- the ground state of nuclear spin in a B-site, that
is we have here {\it homonuclear system of spins at two distinct ground
states}. Each nuclear spin in A-site of this scheme, has {\it two
}internal eigenstates --- ground $|\uparrow \rangle $ and excited $|\Downarrow \rangle $ and in B-site,
accordingly, --- $|\downarrow \rangle $ and $|\Uparrow \rangle $. We take into account that the life
time (the longitudinal or spin-lattice relaxation time $T_{1})$ of
excited states at low temperatures is very long. Each qubit of
quantum information in this state will be represented here,
similar to \cite{3}, by the {\it four} consecutive cells: the logical qubit
basis state "0" will be represented by unit $|\Downarrow  \Uparrow  \uparrow  \downarrow \rangle $, whilst the
state "1" --- by $|\uparrow  \downarrow  \Downarrow  \Uparrow \rangle $. It is important here that the resonant
frequencies of nuclear spins depend on neighbor spins states.\par
The input and output of the information in the array of ground
states spins could be performed at the ends of the array, where
the nuclear spin (say in A-site at the left end) has only one
neighbor spin and distinguishing resonant frequency $\nu _{\mathrm{A},-1/2}$
$(m_{<} + m_{>} = -1/2)$. The corresponding selective resonance RF
$\pi _{\mathrm{A},-1/2}-$pulse inverts only one nuclear spin (in A-site) at the end
of array and doesn't influence on any ones. Then the new selective
RF $\pi _{\mathrm{B},0}-$pulse will invert next nuclear spin (in B-site), which has
the opposite orientation of ground and exited neighbor nuclear
spins ($m_{<} + m_{>} = 0$ in A-site) and consequently the new resonant
frequency, distinguished from the frequency of spins with the
neighbor nuclear spin in ground states $(m_{<} + m_{>} = 1)$. Thus the
qubit state "0", that is $|\Downarrow  \Uparrow  \uparrow  \downarrow \rangle $, is formed in the following way
(the pulses act on underlined spins):
$$
               \underuparrow             \downarrowB        \uparrowA     \downarrowB                \uparrow \ldots \;\buildrel {\pi_{\mathrm{A},-1/2}-\mathrm{pulse}}\over\Rightarrow\;
               \DownarrowA               \underdownarrow    \uparrowA     \downarrowB                \uparrow \ldots \;\buildrel {\pi_{\mathrm{B},0}-\mathrm{pulse}}\over\Rightarrow\;
\mathrel{\mathop{\doubleline{\DownarrowA               \UparrowB          \uparrowA     \downarrowB}}\limits^{"0"}}\uparrow \ldots
$$

The qubit state "1" at the edge of array is formed by means of
still three pulses: at first $\pi _{\mathrm{A},0}$, then $\pi _{\mathrm{A},-1/2}$ and $\pi _{\mathrm{B},0}-$pulses:\par
\par
$$
   \doubleline{\DownarrowA                   \UparrowB      \underuparrow       \downarrowB}        \uparrow  \ldots \;\buildrel {\pi_{\mathrm{A},0}-\mathrm{pulse}}\over\Rightarrow \;
               \underDownarrow     \overline{\Uparrow       \underDownarrow     \underdownarrow     \uparrow} \ldots \;\buildrel {\pi_{\mathrm{A},-1/2}-\mathrm{pulse}}\over\Rightarrow \;
               \uparrow            \overline{\underUparrow  \Downarrow          \underdownarrow     \uparrow} \ldots
$$
$$
\;\buildrel {\pi_{\mathrm{B},0}-\mathrm{pulse}}\over\Rightarrow \;
\mathrel{\mathop{\doubleline{\uparrowA                \downarrowB         \DownarrowA    \UparrowB}}\limits^{"1"}}\uparrow \ldots
$$
\par
The states
$|\overline{\Uparrow \Downarrow \downarrow \uparrow}\rangle$ and
$|\overline{\downarrow \uparrow \Uparrow \Downarrow}\rangle$
may be called as the
reversed states relative to the states
$|\doubleline{\uparrow \downarrow \Downarrow \Uparrow}\rangle$ and
$|\doubleline{\Downarrow \Uparrow \uparrow \downarrow}\rangle$.\par
Note that a random inversion of only one spin will result in
completely destruction of the qubit. 
However, to form, for example, the
error of "0" $\Rightarrow "1"$ type in the coding of stored quantum
information it is essential to invert simultaneously four spins.
Therefore, it may be concluded that the considered way of qubit
coding ensures a better fault-tolerance with respect to this type
of errors.\par
Authors of \cite{8} have considered also another scheme of the
four-spin encoding two logical qubits, which are represented by
the two zero-total states of four spins, generated by the pairs
respectively of the singlet and triplet states. This scheme leads
in the collective decoherence conditions to the fault-tolerant
implementation of quantum computations. The collective decoherence
conditions can be attained in coupled spins at very low
temperatures, where all collective but the longest wavelength
acoustic phonon modes are quenched.\par
The further shift-loading of qubit states into the array is
implemented by means of pulse sequence
$\pi _{\mathrm{A},0}$, $\pi _{\mathrm{B},0}$, $\pi _{\mathrm{A},0}$, $\pi _{\mathrm{B},0}\ldots$,
which is represented by following SWAP operation:\par
$$
\mathrel{\mathop{\doubleline{\uparrowA \downarrowB \underDownarrow \UparrowB}}\limits^{"1"}}\uparrow
\ldots \;\buildrel {\pi_{\mathrm{A},0}-\mathrm{pulse}}\over\Rightarrow\;
   \uparrow \overline{\downarrow\uparrow\Uparrow \underDownarrow}\downarrow
\ldots \;\buildrel {\pi_{\mathrm{B},0}-\mathrm{pulse}}\over\Rightarrow\;
   \uparrow \downarrow \doubleline{\uparrow \downarrow \underDownarrow \Uparrow}
\ldots \;\buildrel {\pi_{\mathrm{A},0}-\mathrm{pulse}}\over\Rightarrow\;
$$
$$
\Rightarrow\;
   \uparrow  \downarrow  \uparrow  \overline{\downarrow \uparrow \underUparrow \Downarrow} \underdownarrow
\ldots \;\buildrel {\pi_{\mathrm{B},0}-\mathrm{pulse}}\over\Rightarrow\;
   \uparrow  \downarrow  \uparrow  \downarrow \mathrel{\mathop{\doubleline{\uparrowA \mathrel{\mathop{\downarrow}\limits_{\mathrm{B}} \DownarrowA \UparrowB}}}\limits^{"1"}} \uparrow  \downarrow  \uparrow  \downarrow \ldots
$$
and so on.\par
The role of the atoms at array ends can play here, as it was
discussed in \cite{2}, also dopand nuclei D at the certain place of the
array with distinct resonant frequency or a defect that modifies
the resonant frequency of the nearest nuclear spin in the array.
Starting from the perfectly initialized states inputting the
information can be performed by setting the dopant D-spin to
desired state by means of pulse at his resonant frequency. The
nuclear spin state of cell nearest to the dopand is created by
SWAP operation mentioned above. After the required information is
loaded, D-spin is reset to the ground state $|0\rangle $. Upon completion
of computation, the state of any qubits can be measured by moving
it to the A-site nearest to D, then swapping $A\Leftrightarrow \mathrm{D}$ and finally
measuring the state of D-spin.\par
\par
\section{One-qubit operations in one dimension}
\par
As in \cite{3}, we introduce still $\pi _{\mathrm{A},1}$ and $\pi _{\mathrm{B},-1}-$pulses and
operators U$_{\mathrm{A},1}$ and U$_{\mathrm{B},-1}$. The last means, that each spin in A- and
B-site is subjected to a unitary transform U, which acts on spins
in A- and B-sites with resonant frequency corresponding to the
both neighbor nuclear spin in the {\it same excited} states
($m_{<} + m_{>} = \pm 1$, sign "$+$" is for excited neighbor spins in B-sites
$|\Uparrow \rangle $, and sign "$-$" --- for A-sites $|\Downarrow \rangle $, see Table). Note, that when
operator $\mathrm{U}$ is a simple inversion, the actions of U$_{\mathrm{A},1}$ and U$_{\mathrm{B},-1}$
are identical to $\pi _{\mathrm{A},1}$ and $\pi _{\mathrm{B},-1}-$pulse.\par
\par
$\;$
\par
{\bf Table.} The $\pi -$pulses for spins in A- and B-sites\par
\par
$$
\begin{tabular}{l c c c c c}
 Neighbor spin states. A-site  &  $\uparrowA \downarrow    $&$ \downarrow \uparrowA \downarrow $&$ \downarrow \uparrowA \Uparrow   $&$ \Uparrow \uparrowA \downarrow $&$ \Uparrow \uparrowA \Uparrow  $ \\
 Resonance frequency           &  $\nu _{\mathrm{A}}(-1/2) $&$ \nu _{\mathrm{A}}(-1)           $&$ \nu _{\mathrm{A}}(0)            $&$ \nu _{\mathrm{A}}(0)          $&$ \nu _{\mathrm{A}}(1)         $ \\
 $\pi -$pulses                 &  $\pi _{\mathrm{A},-1/2}  $&$ \pi _{\mathrm{A},-1}            $&$ \pi _{\mathrm{A},0}             $&$ \pi _{\mathrm{A},0}           $&$ \pi _{\mathrm{A},1}          $
\end{tabular}
$$
\par
$$
\begin{tabular}{l        c c c c c}
 Neighbor spin states. B-site  &  $\downarrowB \uparrow    $&$ \uparrow \downarrowB \uparrow   $&$ \uparrow \downarrowB \Downarrow $&$ \Downarrow \downarrowB \uparrow $&$ \Downarrow \downarrowB \Downarrow  $  \\
 Resonance frequency           &  $\;\nu _{\mathrm{B}}(1/2)  $&$ \;\nu _{\mathrm{B}}(1)\;      $&$ \nu _{\mathrm{B}}(0)            $&$ \nu _{\mathrm{B}}(0)            $&$ \nu _{\mathrm{B}}(-1)       $ \\
 $\pi -$pulses                 &  $\pi _{\mathrm{B},1/2}   $&$ \pi _{\mathrm{B},1}             $&$ \pi _{\mathrm{B},0}             $&$ \pi _{\mathrm{B},0}             $&$ \pi _{\mathrm{B},-1}        $
\end{tabular}
$$
\par
$\;$
\par
Let us construct now the logical gates of quantum computer. At
first we shall investigate at first the scheme for one-qubit gate.
The considered section of the array (Fig.\ 1) contains three qubits
in states "1", "0" and "1", each being separated by number
multiple four of spacer cells -nuclear spins in ground state.
Therefore each qubit requires a total of eight physical spins in
the array (four for the encoding plus four spacers) (Fig.\ 1). As
it was made in \cite{1,3}, we will also introduce here {\it one control unit}, 
(not to be confused with {\it control qubit} in case of CNOT gates),
which is represented here by {\it six} consecutive cells in the pattern
$\doubleline{\Uparrow \Downarrow \downarrow \uparrow \Uparrow \Downarrow}$. The control unit (CU) exists only in one place along
the array and is separated by odd number of spacer cells --- spins
(the scheme at Fig.\ 1 has three).\par
$\;$
\par
\epsfbox{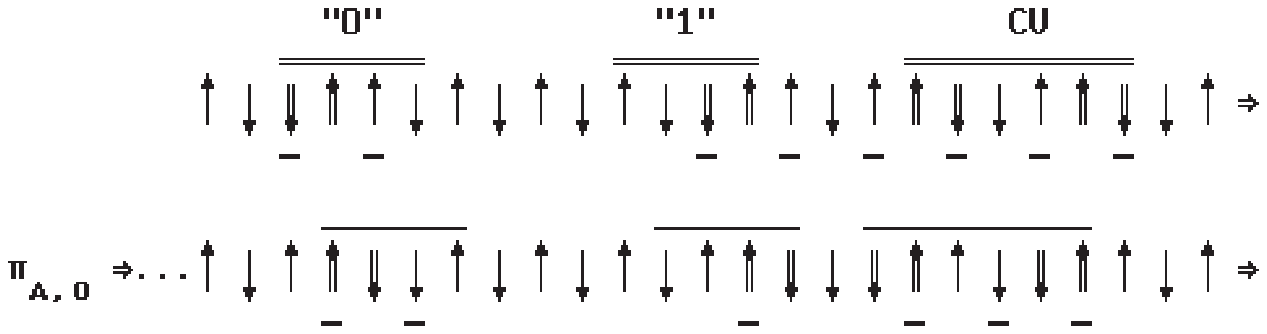}\par
\epsfbox{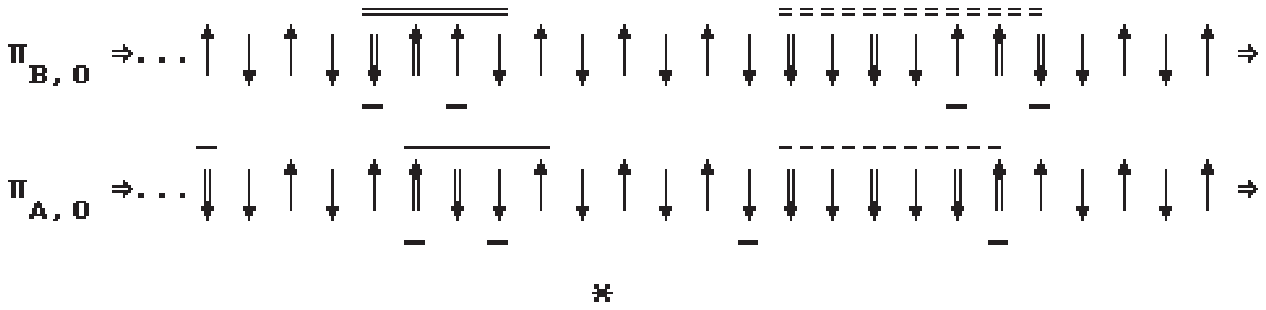}\par
\epsfbox{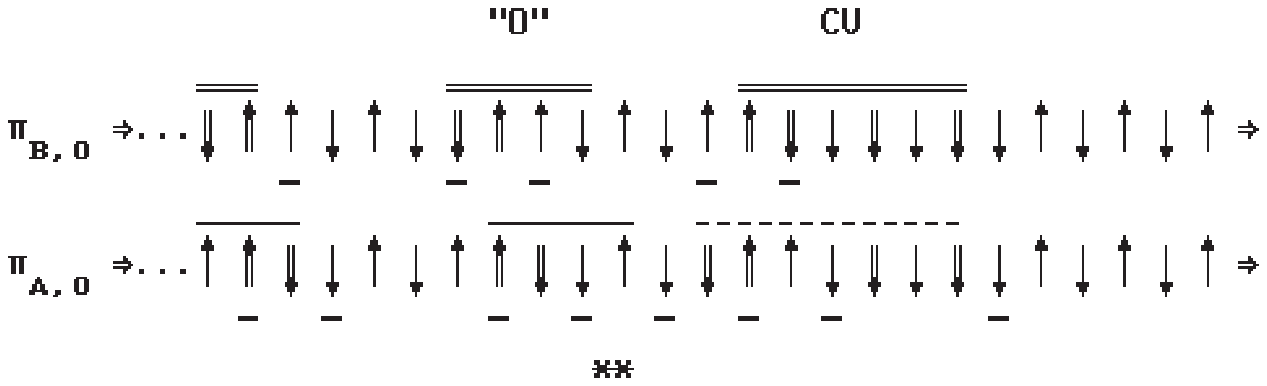}\par
\epsfbox{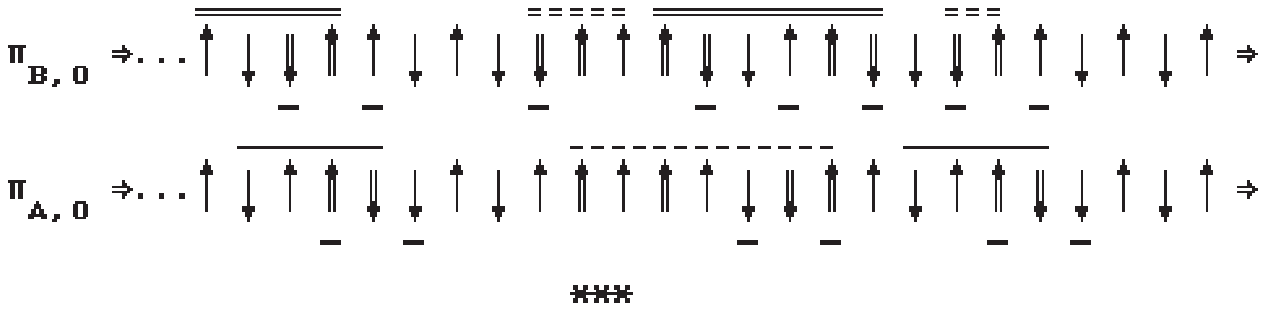}\par
\epsfbox{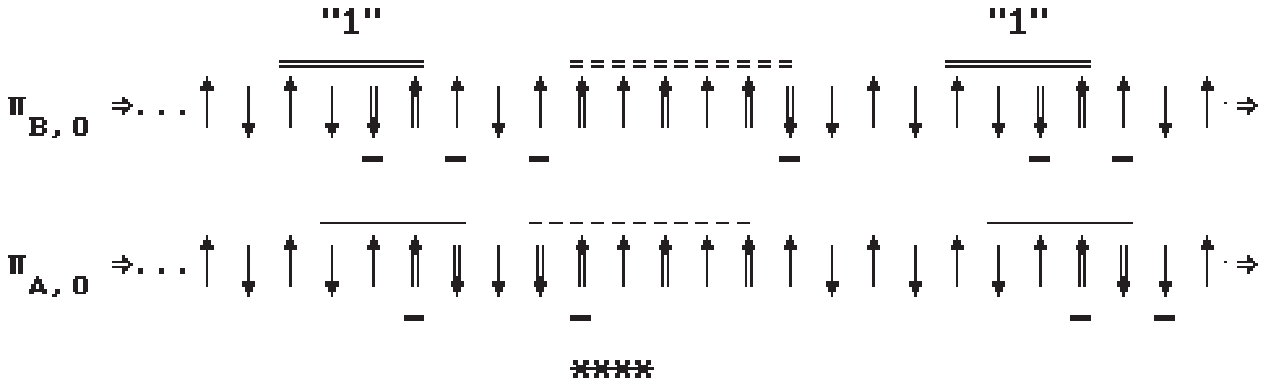}\par
\epsfbox{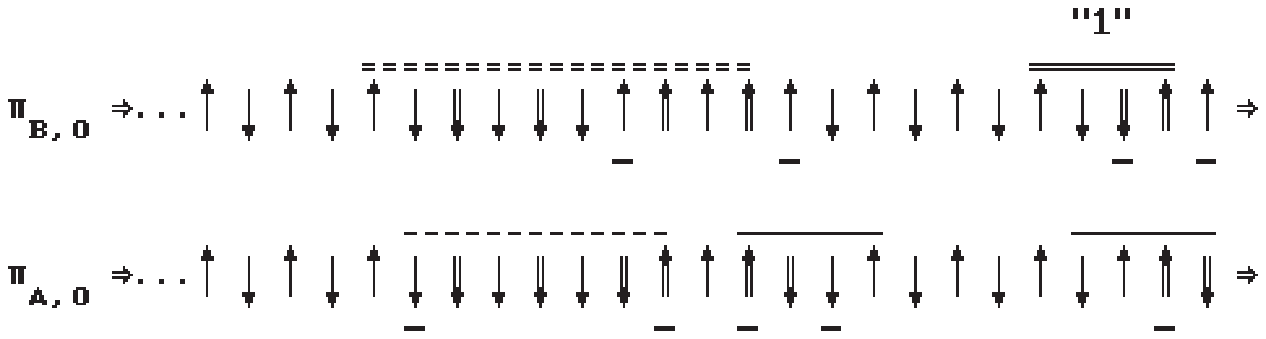}\par
\epsfbox{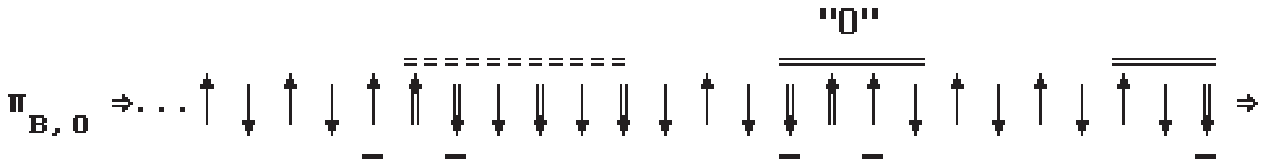}\par
\epsfbox{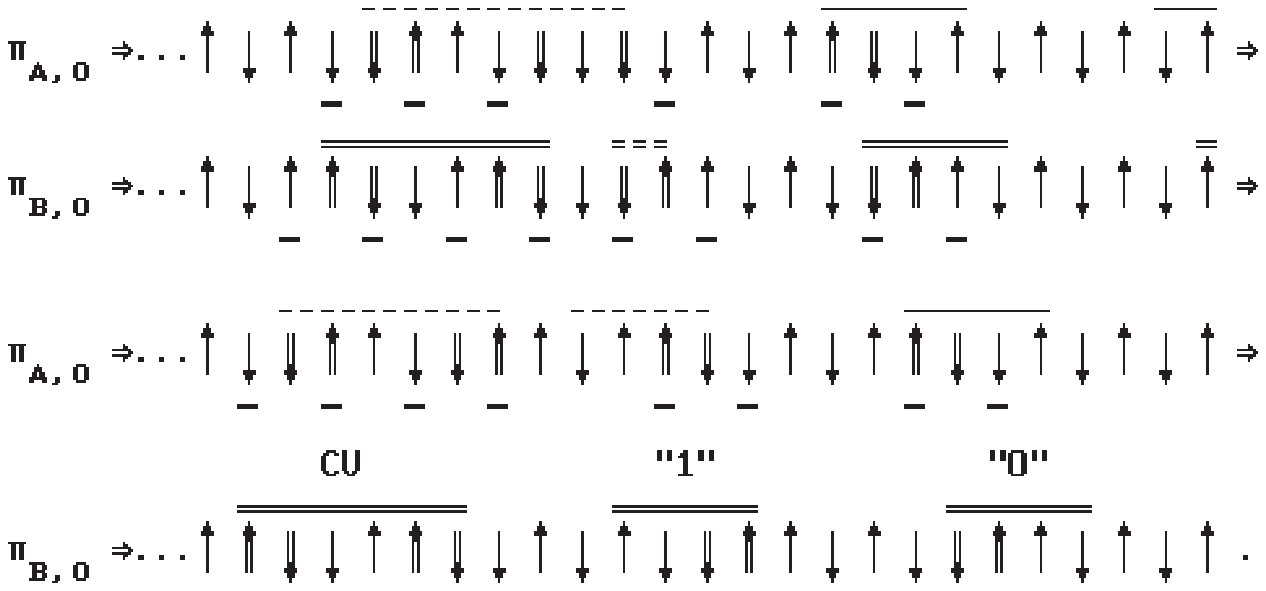}\par
\nobreak\par\nobreak
Fig.\ 1. The scheme of the SWAP update pulse sequence.\par
$\;$
\par
The applying SWAP update sequence of pulses
$\pi _{\mathrm{A},0}$, $\pi _{\mathrm{B},0}$, $\pi _{\mathrm{A},0}$,
$\pi _{\mathrm{B},0}$, $\pi _{\mathrm{A},0}\ldots$
moves the qubits to the right and CU to the left
relative to the qubits, yet the form of the qubits and the CU are
{\it preserved.} The CU passes through qubit in state "1" and "0",
leaving it unchanged and continues further (Fig.\ 1).\par
To implement the one-qubit logical gate the additional
computing updates six pulse sequence
$\pi _{\mathrm{A}1}$, $\pi _{\mathrm{B}1}$, $\pi _{\mathrm{B}0}$, $\pi _{\mathrm{A}1}$, $\pi _{\mathrm{B}0}$, U$_{\mathrm{A},1}$
is applied when CU reaches the mid-way through passing the qubit $\mathrm{Y}$
("1" or "0", marked * and ** at Fig.\ 1). The effect of the
additional sequence is to apply a unitary transform $\mathrm{U}$ only to the
spin representing the qubit $\mathrm{Y}$: $\mathrm{t} = \mathrm{UY}$ (Fig.\ 2). The scheme of the
additional computing update pulse sequence after stage * is shown
at Fig.\ 2. The scheme of sequence after stage ** is shown in Appendix A1.\par
$\;$
\par
\epsfbox{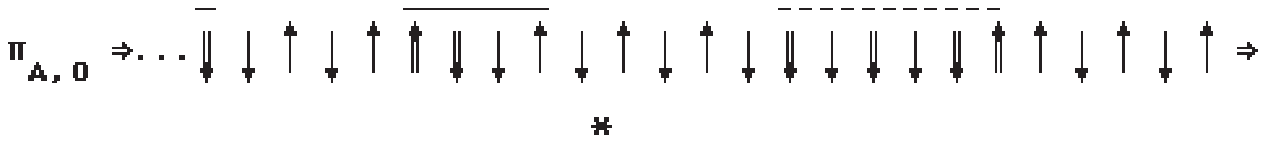}\par
\epsfbox{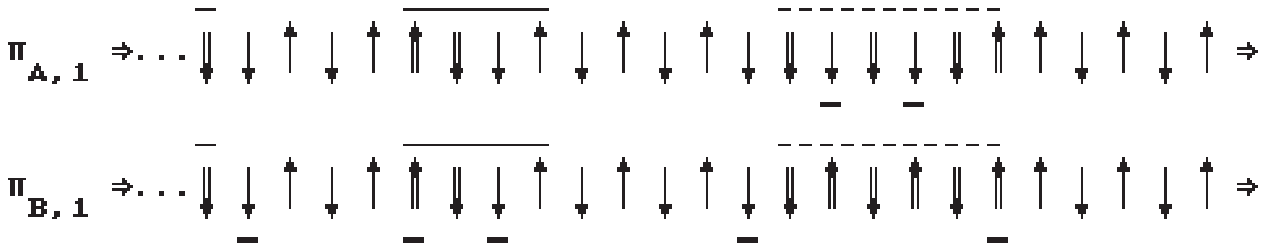}\par
\epsfbox{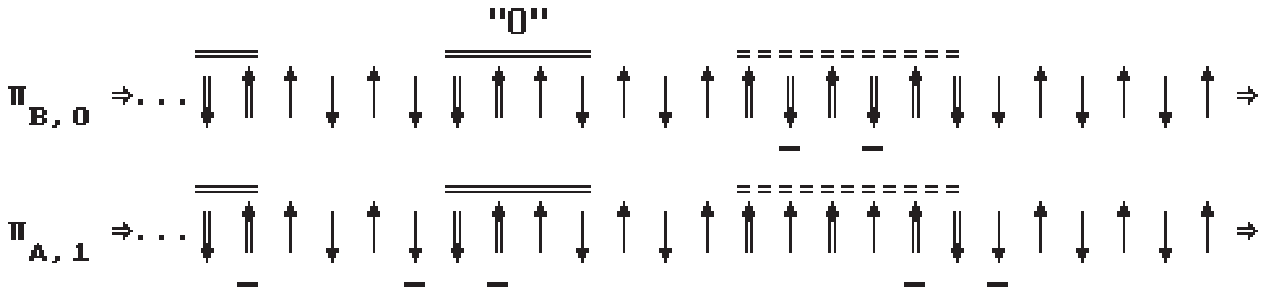}\par
\epsfbox{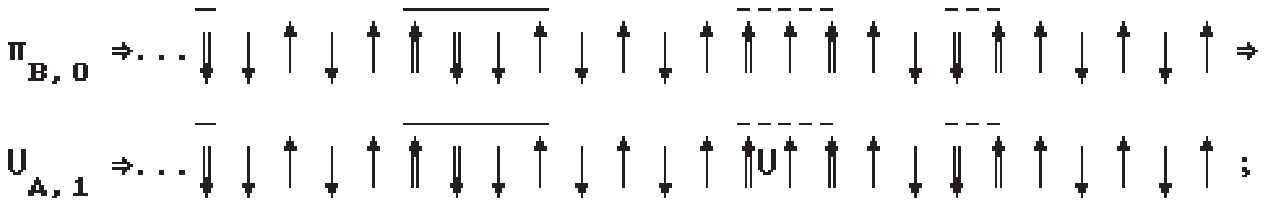}\par
\nobreak\par\nobreak
Fig.\ 2. The scheme of the additional computing update pulse
sequence after $\mathrm{stage\;*}$\par
$\;$
\par
Let us take now the one-cell unitary operation at the end of
additional sequence:
\begin{eqnarray}
\mathrm{U}_{\mathrm{A},1}|\uparrow \rangle  = a|\uparrow \rangle  + b|\Downarrow \rangle  ,  |a|^{2} + |b|^{2} = 1.\label{2}
\end{eqnarray}
\par
Re-applying the update pulses after unitary operation in
reverse order the CU moves away from transformed cell and is
returned to its initial state (see Appendix A2). We will have in
the result the superposition of states:
\begin{eqnarray}
|\psi\rangle
= a|\mathrel{\mathop{\doubleline{\uparrow \downarrow \Downarrow \Uparrow}}\limits^{"1"}}\rangle
+ b|\mathrel{\mathop{\doubleline{\Downarrow \Uparrow \uparrow \downarrow}}\limits^{"0"}}\rangle.\label{3}
\end{eqnarray}
\par
The CU and additional computing updates pulse sequence together
ensure the computing operations with qubits. Note, such one-qubit
gate requires seventeen elementary pulses.\par
Not let us consider the state of quantum register, shown at
the first line on Fig.\ 1 and apply after stage marked ** the pulse
$\pi _{\mathrm{A},1}$. The result is presented on Fig.\ 3.\par
$\;$
\par
\epsfbox{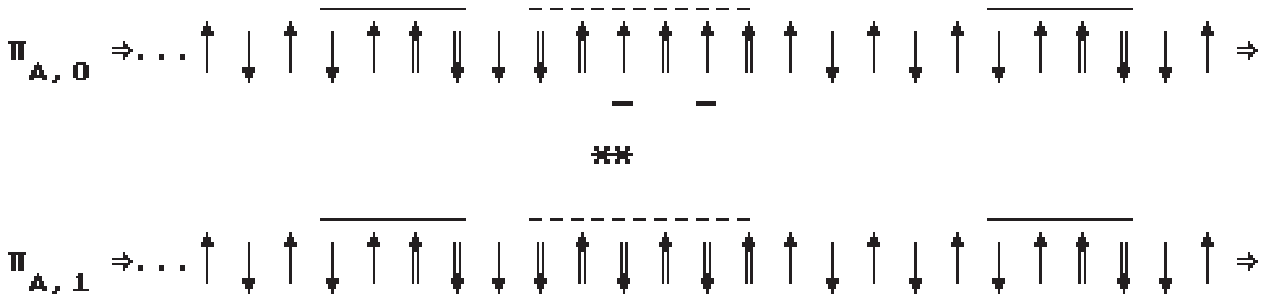}\par
\nobreak\par\nobreak
Fig.\ 3. The result of the SWAP pulse sequence finished by pulse
$\pi _{\mathrm{A},1}$.\par
$\;$
\par
We see that the CU moves transparently past the qubit "1" and
continues until mid-way through passing qubit "0". Now the CU
itself is subject to a transformation: it is altered from
$\doubleline{\Uparrow \Downarrow \downarrow \uparrow \Uparrow \Downarrow}$ to
$\overDownarrow \overUparrow \overDownarrow \overUparrow \overDownarrow \overUparrow$
(only for passing the qubit "0"!) and
qubit "0" itself will be destroyed (Fig.\ 3).\par
Now we will apply again the SWAP pulse sequence and will become:\par
$\;$
\par
\epsfbox{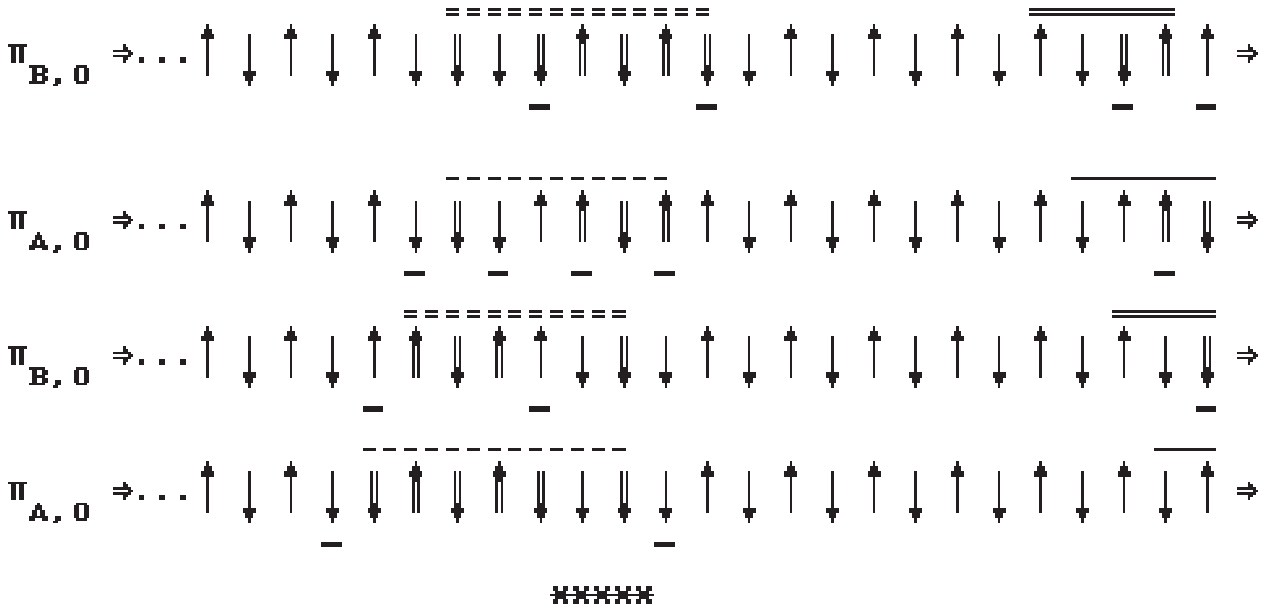}\par
\epsfbox{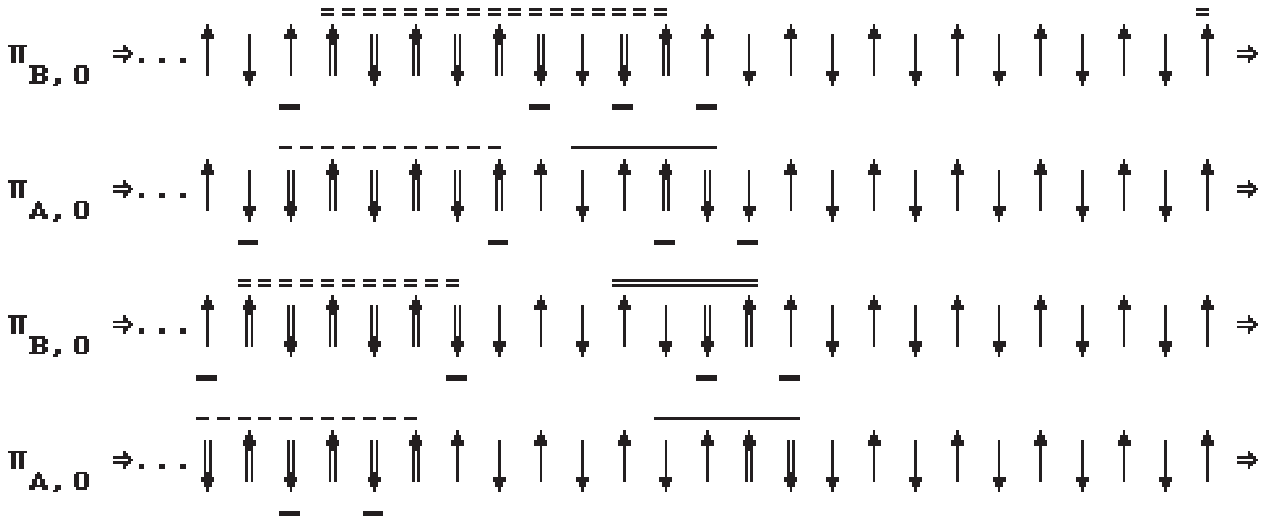}\par
\nobreak\par\nobreak
Fig.\ 4. The scheme of the SWAP pulse sequence following sequence Fig.\ 3.\par
$\;$
\par
We see here, that by passing qubit "1" the altered CU is
preserved his form.\par
\par
\section{Two-qubit operations in one dimension}
\par
To implement the two-qubit gate such as CNOT it should be
applied the another update pulse sequence.\par
Let now the qubit "0" will be as {\it control qubit} of CNOT gate.
The CU transforms in altered form by passing the control qubit and
then we extend the sequence Fig.\ 5 after stage marked ***** by
following pulses with the end inversion pulse U$_{\mathrm{A},1} \equiv \pi _{\mathrm{A},1}$:
\begin{eqnarray}
\pi _{\mathrm{A},1},\; \pi _{\mathrm{B},1},\; \pi _{\mathrm{B},0},\; \pi _{\mathrm{A},0},\; \pi _{\mathrm{A},1},\; \pi _{\mathrm{B},0},\; \pi _{\mathrm{A},0},\; \pi _{\mathrm{B},1},\; \pi _{\mathrm{A},1}.\label{4}
\end{eqnarray}
\par
We will obtain:\par
$\;$
\par
\epsfbox{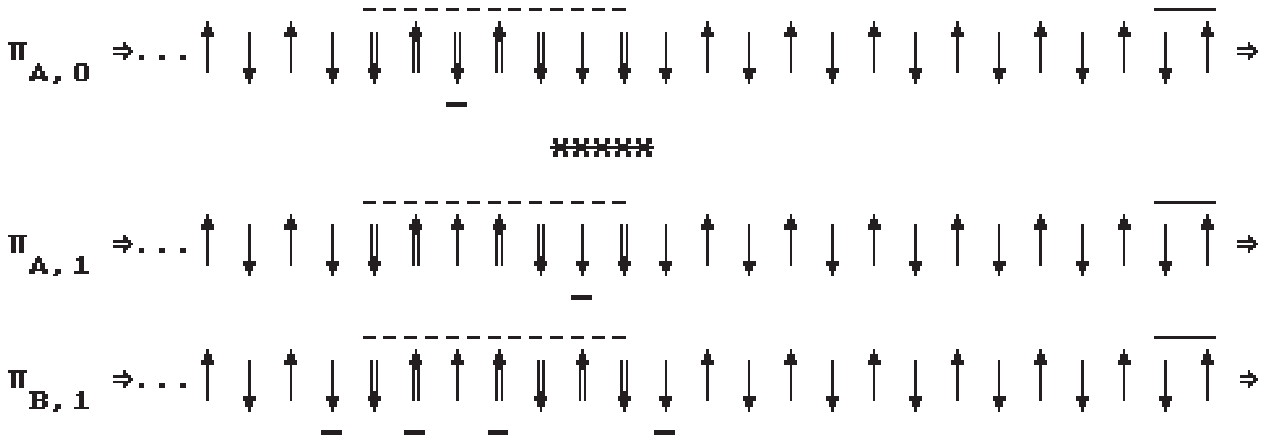}\par
\epsfbox{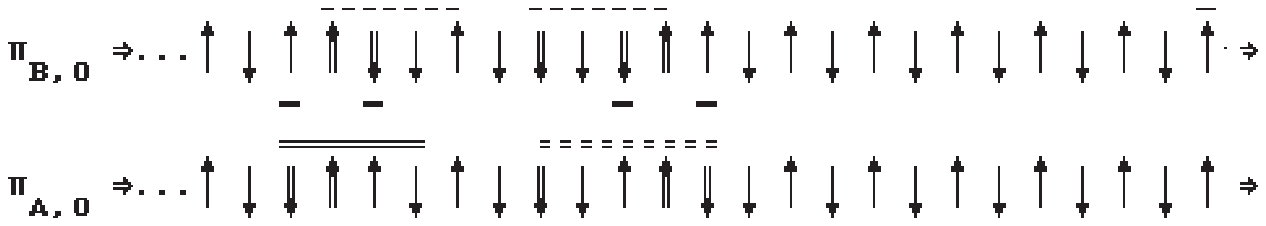}\par
\epsfbox{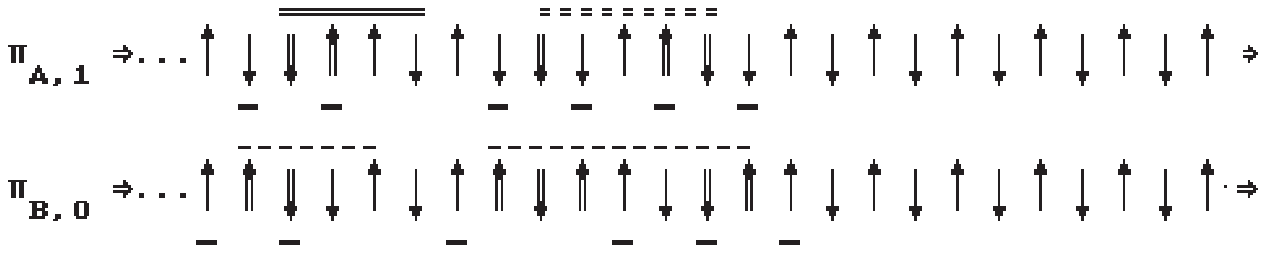}\par
\epsfbox{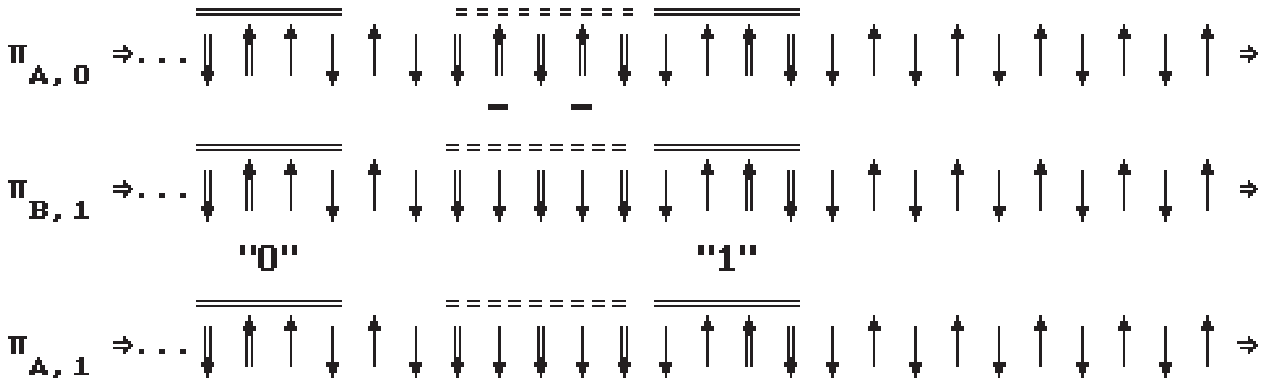}\par
\nobreak\par\nobreak
Fig.\ 5. The scheme of the update pulse sequence (\ref{4})\par
$\;$
\par
The last inversion operation {\it doesn't have effect} on target
qubit "1", as it must be for CNOT gate. The reverse sequence
returns CU and qubits to their initial states.\par
Let us return then to sequence Fig.\ 1 and continue it after
stages marked **, when CU passes mid-way through qubit "1" by
following sequence:\par
$\;$
\par
\epsfbox{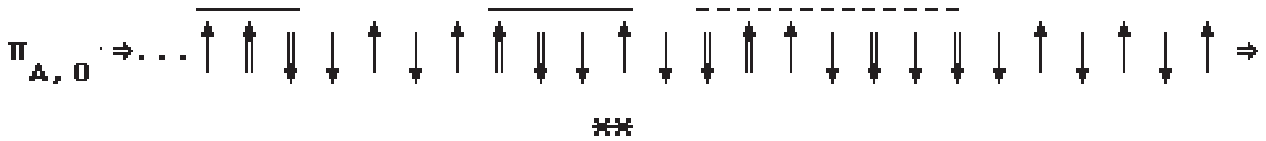}\par
\epsfbox{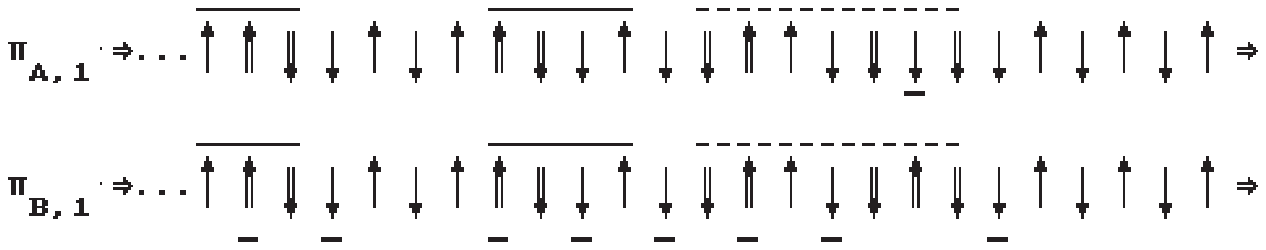}\par
\epsfbox{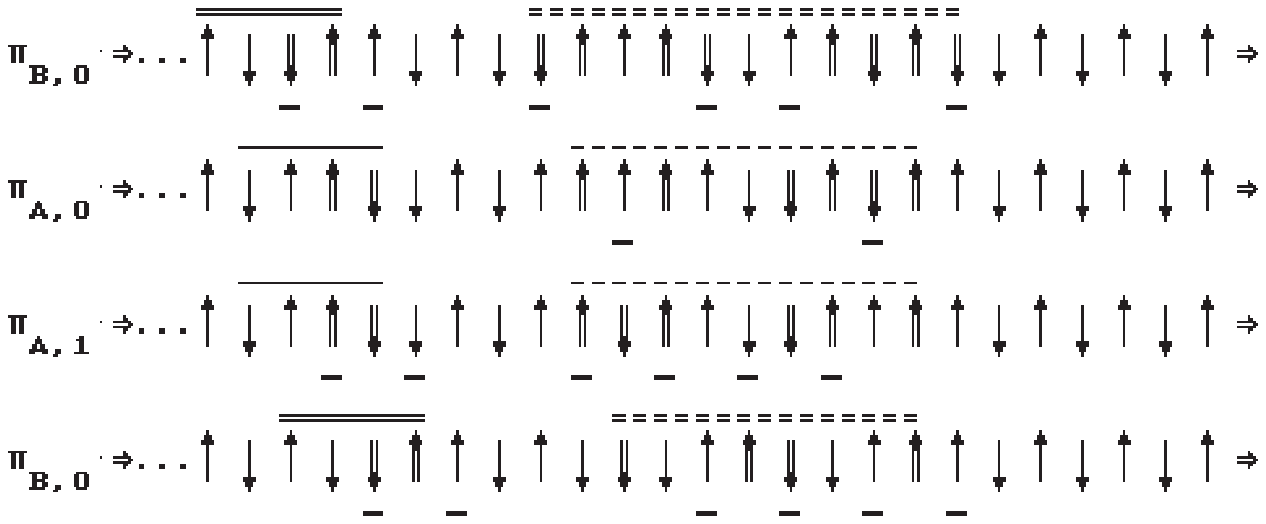}\par
\epsfbox{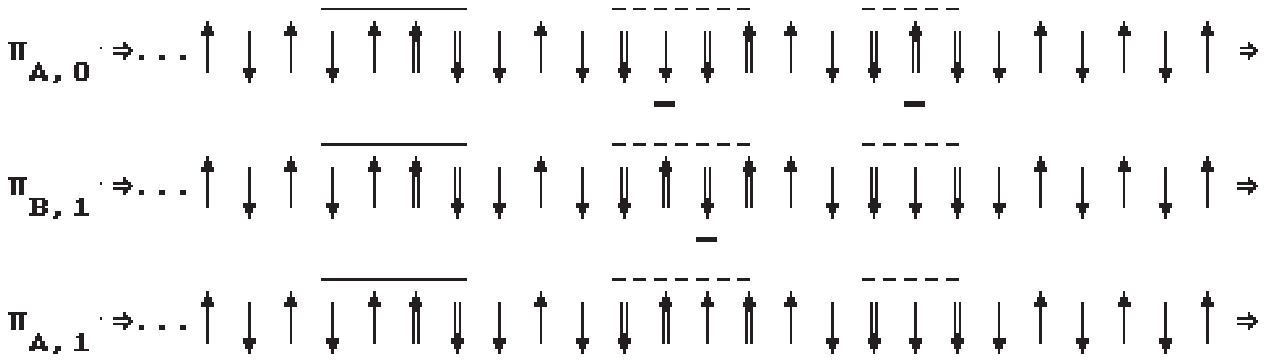}\par
\nobreak\par\nobreak
Fig.\ 6. The scheme of the update pulse sequences after stage **
when CU passes the qubit "1"\par
$\;$
\par
The scheme of the update pulse sequences after stage **** when
CU also passes the qubit "1" is shown in Appendix A3.\par
We see, that the last inversion operation has effect on target
qubit, as it must be for CNOT gate only when the control qubit is
"1".\par
We can consider a large set of quasi-one-dimensional
antiferromagnetically ordered weakly coupled at $J(l_{\mathrm{y}}) \gg  J(l_{\mathrm{x}})$
arrays donors $^{31}P$ in silicon substrate as an ensemble of
artificial molecules. In this case, there is no need to address
qubits individually. We suppose that for the determination of
nuclear spin states in ensemble of those identical artificial
molecules, as in liquid-state bulk-ensemble quantum computer
\cite{8,9}, there are {\it no need to fulfill the electrical measurements
}and consequently {\it any electrodes.} Since the read-out signal in this
case will be proportional to the numbers of artificial molecule in
the ensemble, it may be used the NMR or fluorescence techniques
for ensemble measurement of spin states.\par
\par
\section{Two- and three-dimensional antiferromagnetic structures}
\par
Instead of generalization to the parallel model employing
"sub-computers" with one-dimensional structure, which was
considered in \cite{3}, our approach allows to use also two and
three-dimensional structures. The coupled antiferromagnetically
ordered chains model can be extended to a two-dimensional
antiferromagnetic chess-type ordering. Let the electronic spins of
the neighboring chains are setting for $J(l_{\mathrm{y}}) \neq
J(l_{\mathrm{x}}) > 2\mu _{\mathrm{B}}B$. The electronic spins of
two neighbor chains will be in the singlet ground state. The
subarray of nuclear spins will have the opposite orientation of
nuclear spins relative to the subarray of neighboring chain. The
electron subsystem of two neighbor chains is in antiphase state,
that is have the half period shift of antiferromagnetically
ordered electronic spins in the one chain relative to the other.
The nuclear subsystem of the both chains becomes corresponding
chess-type ordering (Fig.\ 7).\par
Let us suppose that an initial
state containing some number of cells is loaded in the
two-dimensional structure of many coupled chains. The inputting
the information into the cell nearest to dopant atoms D (D-spin)
is performed by means of corresponding $\pi -$pulse: $\downarrow
\Rightarrow  \Uparrow$ or $\uparrow \Rightarrow \Downarrow$.
Then, the resonant frequencies of neighbor spins in the same chain
and of spins in neighbor chain is altered and another $\pi -$pulse
will invert one of they or both according to the values $I_{1}$
and $I_{2}$ of indirect nuclear spins interaction inside and
between the chains. Therefore, the excited nuclear spin states and
accordingly the qubits may be passed to any place in all
two-dimensional structure. We will suppose as before that the
qubits state will be represented by four spin states in chains.
The computing operation can be fulfilled analogous to the
above-considered one-dimensional scheme.\par 
$\;$
\par
\epsfbox{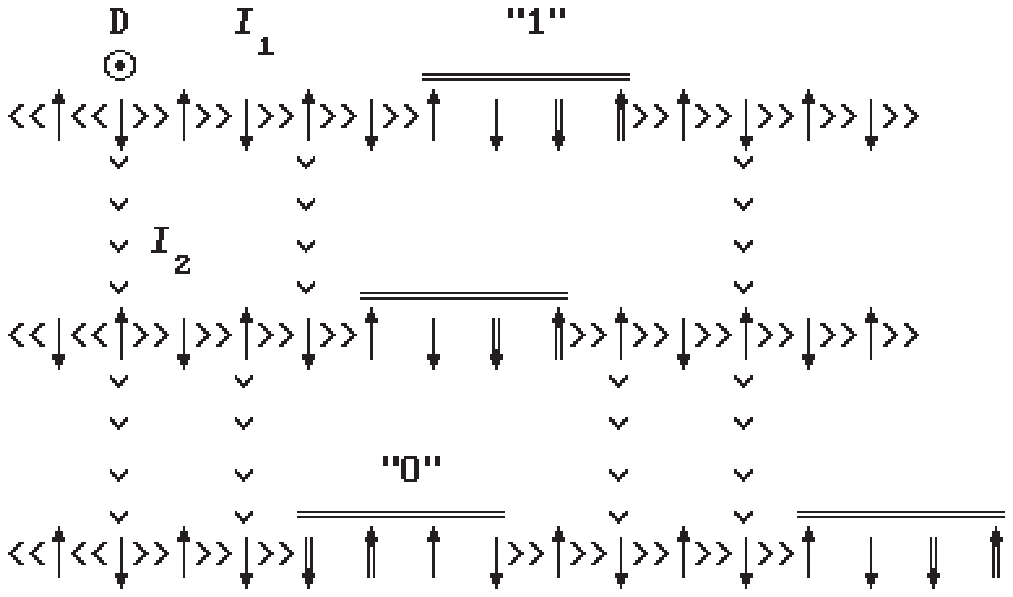}\par
\nobreak\par\nobreak
Fig.\ 7. The scheme of two-dimensional nuclear spin ordering in
antiferromagnetic structure. It is showed the different ways that
connect the D-spin (marked $\odot$) and a certain qubit.\par
$\;$
\par
For everyone CU we have a two-dimensional section of array with
{\it large enough number of qubits} and one dopant atom. It is
defined as a single-domain antiferromagnetic sample. There are
many ways that connect the D-spin and the qubit (Fig.\ 7). This
section plays role of many-qubit artificial molecule and the whole
structure represent a {\it large ensemble} of such molecules, which
work simultaneously and ensure the parallelism of quantum
operations.\par The structures with two and three-dimensional
antiferromagnetic and ferrimagnetic order may be found perhaps
among the {\it natural rare earth} or {\it transition element}
dielectric compounds.\par
The electron magnetization of a one
subarray of antiferromagnet is defined by expression
\cite{13}:
\begin{eqnarray} 2\mu _{\mathrm{B}}N\langle
S_{\mathrm{j}\mathrm{z}}\rangle  = 2\mu _{\mathrm{B}}N (1 - P(T) -
\psi )/2 ,\label{5}
\end{eqnarray}
where for low temperatures in {\it spin-wave approximation}
\begin{eqnarray}
P(T) = \frac{1}{(2\pi )^{\mathrm{d}}} \int \frac{d^{\mathrm{d}}k}{\exp (\epsilon (k)/kT - 1} \label{6}
\end{eqnarray}
is the contribution of thermal and $\psi $  --- of quantum fluctuations,
$\epsilon (k) = \sqrt{\epsilon ^{2}_{0} + (Jak)^{2}}$ --- spin-wave spectrum, $a$ --- lattice period, Z ---
number of near neighbors, $d$ --- dimension of structure.\par
For antiferromagnetic state of {\it easy-axis} type, when the
interaction Hamiltonian of two electron spin $j$ and $g$ for single-axis crystal has the form:
\begin{eqnarray}
H_{\mathrm{j},\mathrm{g}} = J \mathbf{S}_{\mathrm{j}}\mathbf{S}_{\mathrm{g}} - J_{\mathrm{A}}(S_{\mathrm{j}\mathrm{x}}S_{\mathrm{g}\mathrm{x}} + S_{\mathrm{j}\mathrm{y}}S_{\mathrm{g}\mathrm{y}}) ,\label{7}
\end{eqnarray}
where $J > J_{\mathrm{A}} > 0$, $J_{\mathrm{A}}$ --- anisotropy
constant, the contribution of quantum fluctuation, as shown in
\cite{13}, $\psi  = 0$. In addition, for $k \Rightarrow  0$, $T
\ll  \epsilon _{0}$, where  $\epsilon _{0} \sim
\,\mathrm{Z}\sqrt{JJ_{\mathrm{A}}}$,
\begin{eqnarray}
P(T) \sim \mathrm{Const}\cdot (kT\epsilon
_{0}/J^{2})^{\mathrm{d}/2} \exp (-\epsilon _{0}/kT) \Rightarrow  0,\label{8}
\end{eqnarray}
that is in this state the thermal fluctuations in electron system
also are {\it not essential} for the ground electron spin states.\par
Note, that in the case of easy-flat state, $\psi  \neq  0$, but the NMR
resonance frequency depends on the neighbor nuclear states to a
greater extent than in case of easy-axis state \cite{14}.\par
The quantum state {\it decoherence} at low temperatures is defined
on the one hand by the active role of electron spin-wave effects
\cite{14}. They generate the fluctuated local field due to Raman
process of the electron spin wave scattering on individual nuclear
spins. The decoherence time or transverse relaxation time $T_{2}$ of
NMR in antiferromagnet for the low temperatures $(\epsilon _{0}/kT \gg  1)$ then
is determined by expression \cite{14}
\begin{eqnarray}
1/T_{2} \sim  (A^{2}/J)\cdot (kT/J)^{3} (\epsilon _{0}/kT) \exp (-\epsilon _{0}/kT)/\pi ^{2}\hbar  \Rightarrow  0 ,\label{9}
\end{eqnarray}
value $T_{2}$ {\it rapidly grows}.\par On the other hand
decoherence is defined by inhomogeneity of the local magnetic
fields and spread in resonance frequencies. The nuclear spin-spin
interaction in natural dielectric antiferromagnets is defined
mainly by the Suhl-Nakamura indirect mechanism of interaction
through exchange of spin waves and is typically greater than the
value, determined by the direct nuclear spin-spin dipole
interaction. This interaction of nuclear spins could play a large
role in the case of high spin concentration. Both of these
decoherence mechanisms can be, in principle, suppressed by some
NMR many-pulse methods using the stroboscopic observation of spin
dynamics \cite{15,16}.\par The general requirements for natural
antiferromagnetic structures, required for the construction of NMR
quantum computers, can be formulated in the following way:\par 
1) The operating temperature $T$ must correspond to the fully ordered antiferromagnet 
$T_{\mathrm{N}\mathrm{S}} \gg T \gg T_{\mathrm{N}\mathrm{I}}$ 
and to fully polarized nuclear spins
$T_{\mathrm{N}\mathrm{S}}A/J \sim  A/k > T \gg T_{\mathrm{N}\mathrm{I}}$. 
From where we will have the value $T \geq  10^{-3}\,\mathrm{K}$.\par 
2) The two-dimensional and
tree-dimensional magnetic structure must have chess-type order
(see Fig. 7).\par 
3) The magnetic structure must have the {\it
easy-axis} state of antiferromagnetism in single-axis
crystals.\par 4) The atoms must have nuclear spins $I = 1/2$.
Electron spins may be $S \leq  1/2$.\par There are the rare earth
compounds of unique {\it thulium} stable isotope
$^{169}\mathrm{Tm}$, that has nuclear spin $I = 1/2$,
$g_{\mathrm{N}} = 0.458$ and makes up 100\% of naturally occurring
elements with stable spinless isotopes of other elements. They can
be: $\mathrm{Tm}_{2}\mathrm{O}_{3}$, $\mathrm{TmSi}_{2}$, $\mathrm{TmGe}_{2}$ and $\mathrm{TmSe}$
\cite{11,12}. The ground electronic state of magnetic ions
$\mathrm{Tm}^{3+}$ corresponds to $S = 1$. The natural elements O,
Si, Ge and Se have, accordingly, nuclear spin containing isotopes
(in brackets it is shown the isotope occurrence) $^{17}\mathrm{O}$
$I = 5/2$ (0.04\%), $^{29}\mathrm{Si}$\ $I = 1/2$ (4.7\%),
$^{73}\mathrm{Ge}$\ $I = 9/2$ (7.76\%), $^{77}\mathrm{Se}$\ $I =
1/2$ (7.78\%). For $\mathrm{Yb}_{2}\mathrm{O}_{3}$
$T_{\mathrm{NS}} \sim  2.3\,\mathrm{K}$, isotope
$^{171}\mathrm{Yb}$ (14.31\%) has $I = 1/2$ and $S^{\prime} = 1/2$
(ground state is Kramers doublet). It is known, that compound $\mathrm{TmSe}$
has the critical temperature for antiferromagnetic transition
$T_{\mathrm{N}\mathrm{S}} \sim  2K \cite{12}$. The
antiferromagnets with two different nuclear spin $I = 1/2$, for
example $\mathrm{FeF}_{2}$ with rutile-type and $\mathrm{TmAg}$ with
$\mathrm{CsCl}-$type structure, which have critical temperature 79 K and 9.5
K, may be also of interest to the considered questions. Isotopes
$^{57}\mathrm{Fe}$ (2.19\%), $^{19}\mathrm{F}$ (100\%),
$^{107+109}\mathrm{Ag}$ (100\%) have according values
$g_{\mathrm{N}} = 0.182$, $g_{\mathrm{N}} = 5.26$ and
$g_{\mathrm{N}} = 0.24$.\par In conclusion, we will point out the
several advantages of the considered model: it uses the
antiferromagnetic structure containing only one type of atoms with
nuclear spin 1/2, it is not needed to have any gate electrodes,
the decoherence associated with noise voltage is absent, the
considered way of qubit coding ensures a better fault-tolerance
with respect to the generation of wrong qubits, the model admits
an ensemble address qubits, it may be used as base for development
of bulk-ensembles three-dimensional solid-state NMR quantum
computer.\par The author is grateful to K.A.Valiev for critical
reading of the article and useful remarks and V.A.Kokin for the
help in preparation of this text.\par
\par
\section*{Appendixes}
\par
\subsubsection*{A1. The scheme of the additional computing update pulse sequence
after stage ** at Fig.\ 1:}
\par\nopagebreak
$\;$
\par\nopagebreak
\epsfbox{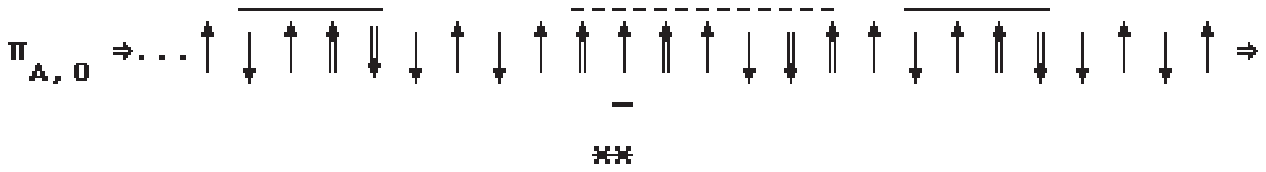}\par
\epsfbox{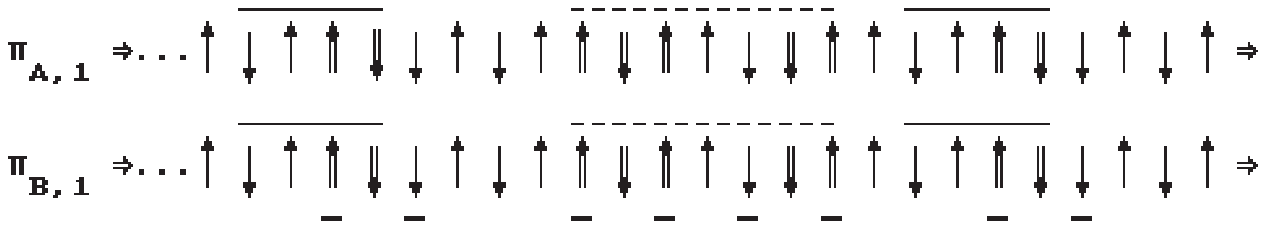}\par
\epsfbox{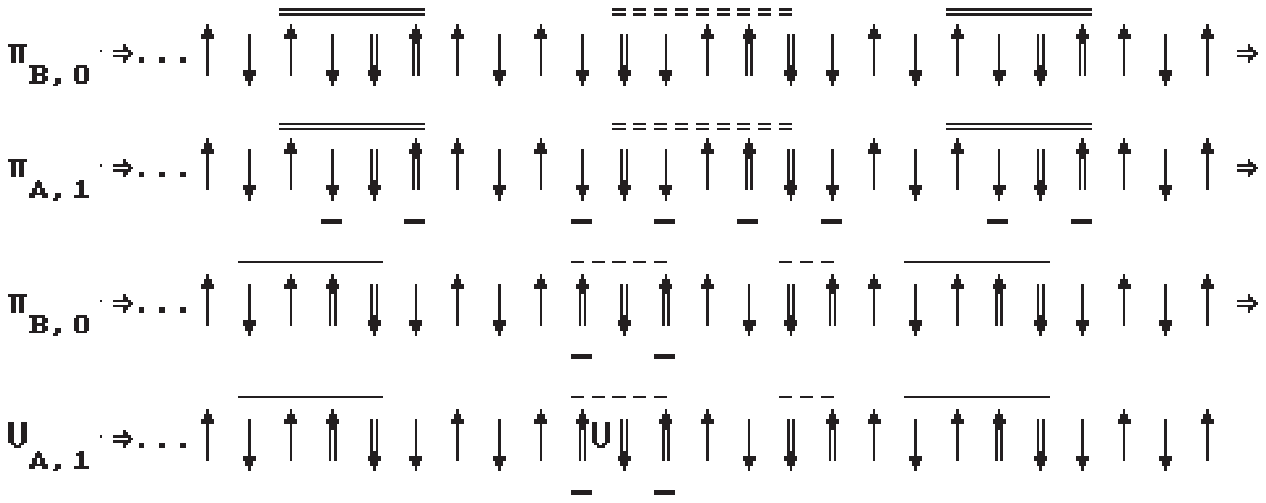}\par
$\;$
\par
\par
\subsubsection*{A2. The scheme of the reverse update pulse sequence after
one-qubit operation U$_{\mathrm{A},1}$:}
\par\nopagebreak
$\;$
\par\nopagebreak
\epsfbox{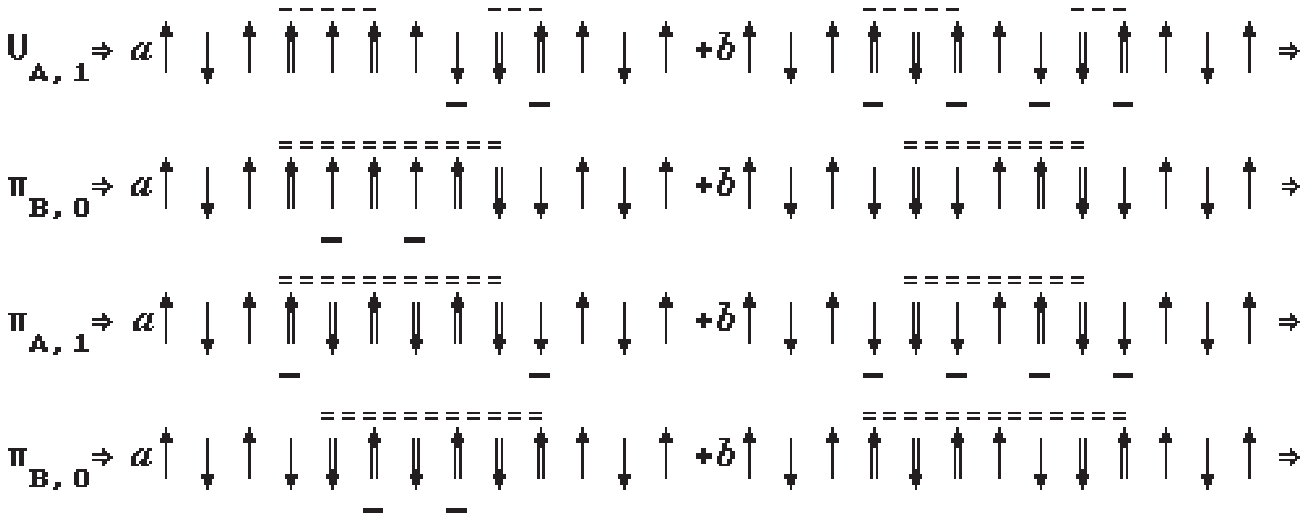}\par
\epsfbox{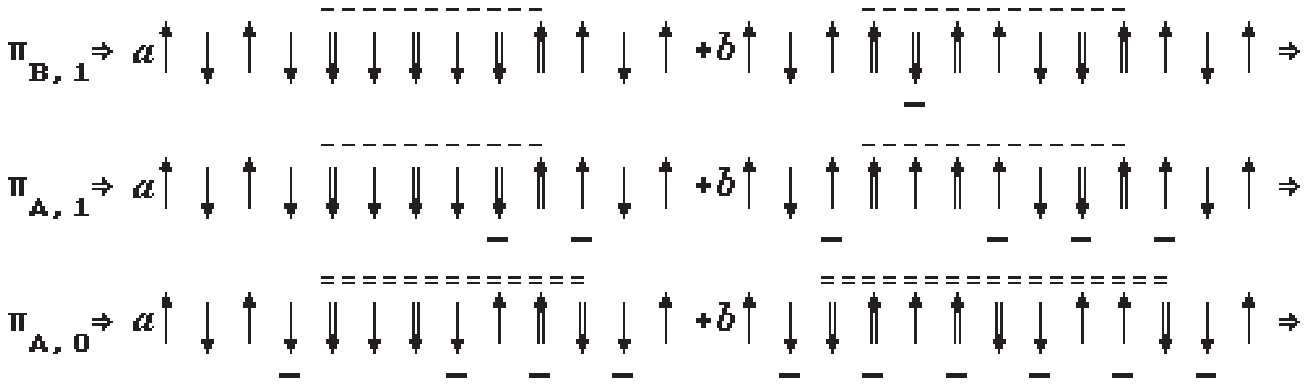}\par
\epsfbox{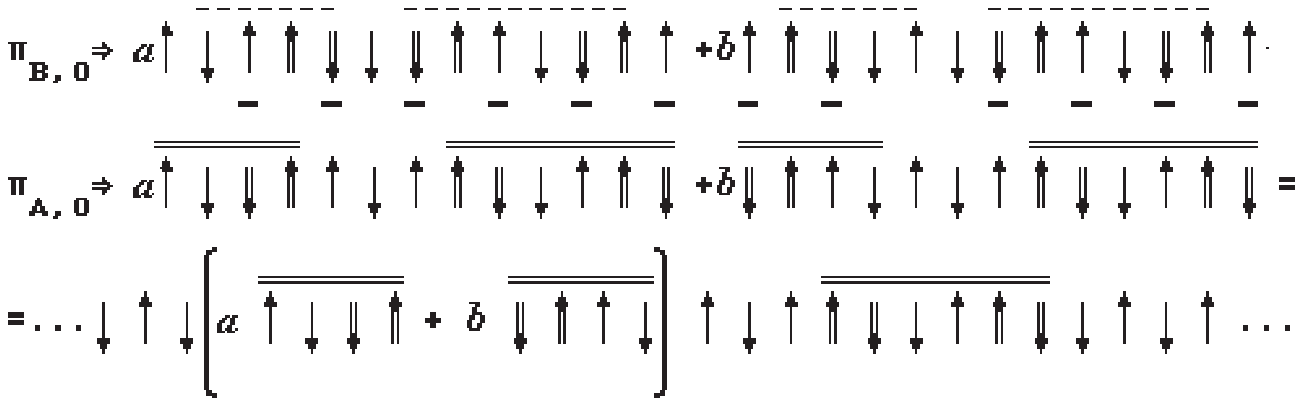}\par
$\;$
\par
\subsubsection*{A3. The scheme of the update pulse sequences when CU passes the
qubit "1" after stage ****:}
\par\nopagebreak
$\;$
\par\nopagebreak
\epsfbox{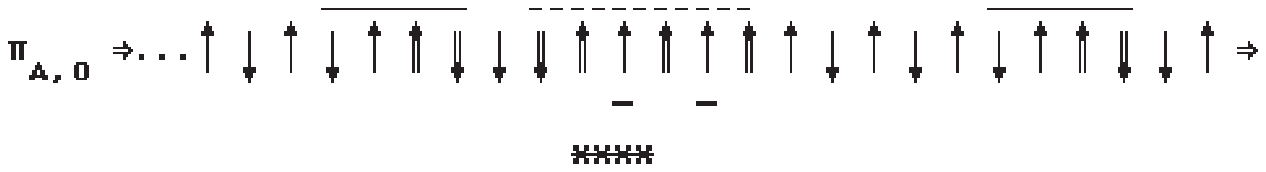}\par
\epsfbox{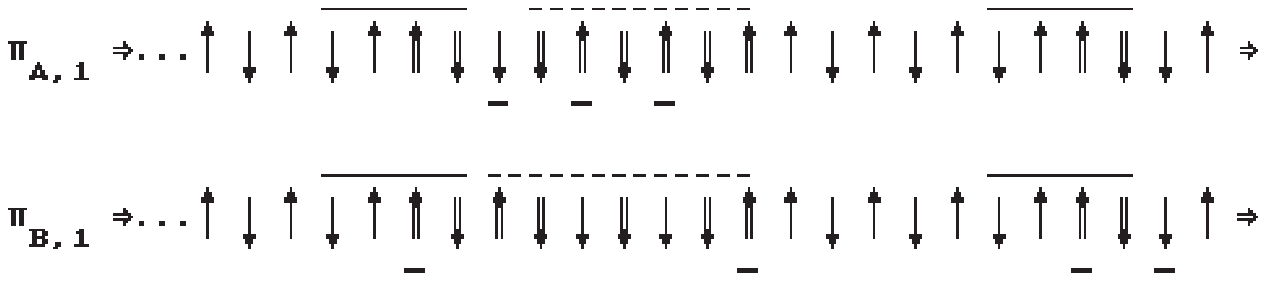}\par
\epsfbox{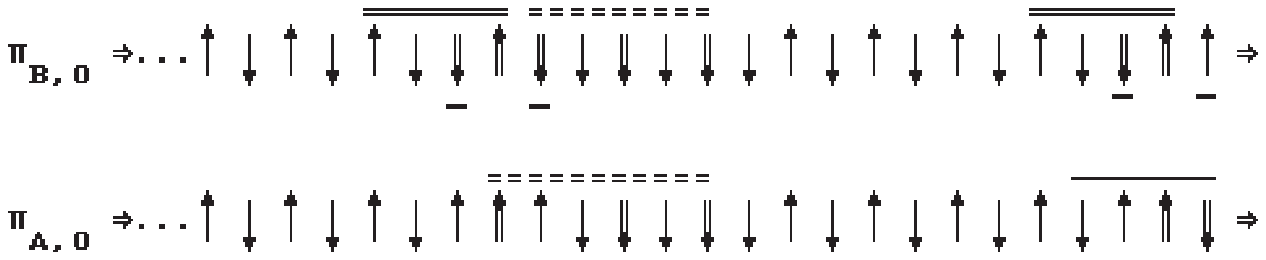}\par
\epsfbox{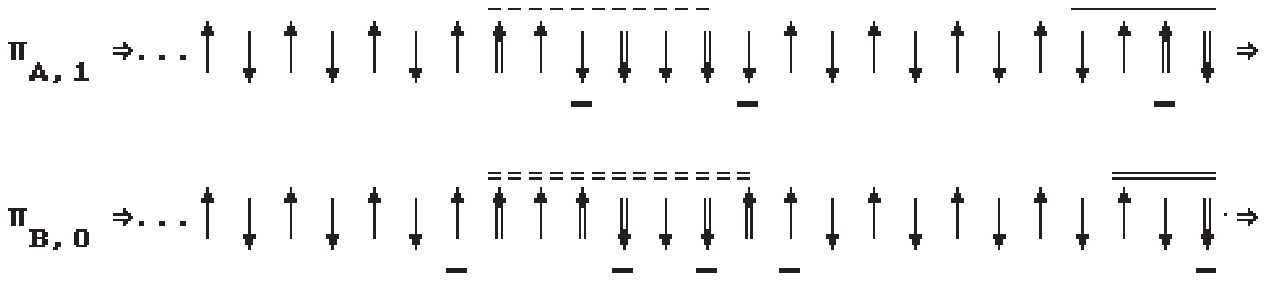}\par
\epsfbox{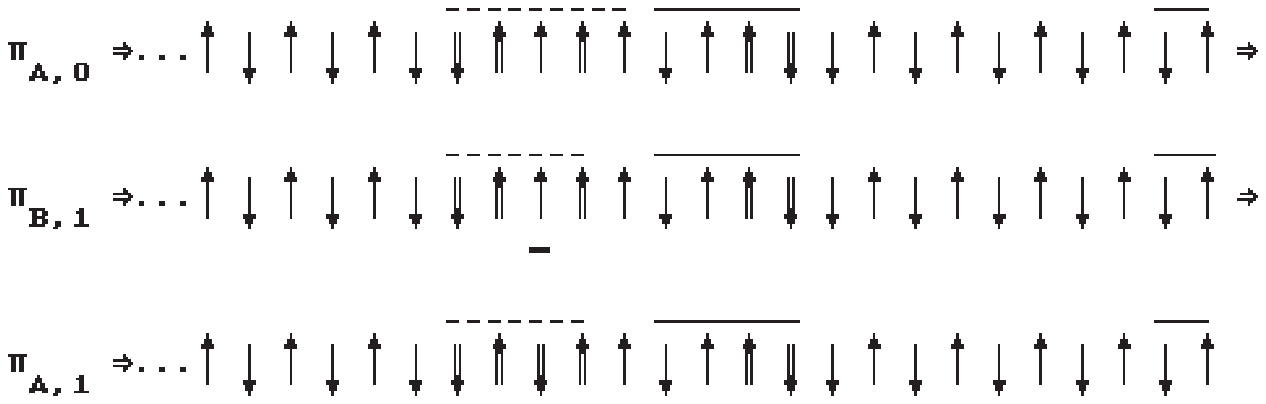}\par
$\;$
\par
\par

\end{document}